\documentclass[12pt,onecolumn,preprint,twocolumn,trackchanges]{aastex6}
\usepackage{natbib}
\usepackage{multirow,color}
\usepackage{bbding}
\usepackage{amssymb}
\usepackage{ulem}
\usepackage{color}
\usepackage{longtable}
\usepackage{array}
\usepackage{bm}
\usepackage{url}
\def\gsim{\;\lower4pt\hbox{${\buildrel\displaystyle >\over\sim}$}\;}
\def\lsim{\;\lower4pt\hbox{${\buildrel\displaystyle <\over\sim}$}\;}
\def\grls{\;\lower4pt\hbox{${\buildrel\displaystyle >\over <}$}\;}

\begin{document}
\title{A comparative study between a failed and a successful eruption initiated from the same polarity inversion line in AR 11387}



\author{Lijuan Liu}
\affiliation{School of Atmospheric Sciences, Sun Yat-sen University, Zhuhai, Guangdong, 519082, China}

\author{Yuming Wang}
\affiliation{CAS Key Laboratory of Geospace Environment, Department of Geophysics and Planetary Sciences, University of Science and Technology of China, Hefei, Anhui, 230026, China}

\author{Zhenjun Zhou}
\affiliation{School of Atmospheric Sciences, Sun Yat-sen University, Zhuhai, Guangdong, 519082, China}

\author{Karin Dissauer}
\affiliation{Institute of Physics, University of Graz, Universit\"atsplatz 5/II, 8010 Graz, Austria}

\author{Manuela Temmer}
\affiliation{Institute of Physics, University of Graz, Universit\"atsplatz 5/II, 8010 Graz, Austria}

\author{Jun Cui}
\affiliation{School of Atmospheric Sciences, Sun Yat-sen University, Zhuhai, Guangdong, 519082, China}
\affiliation{CAS Key Laboratory of Lunar and Deep Space Exploration, National Astronomical Observatories, Chinese Academy of Sciences, Beijing, 100012, China}

\email{ljliu@mail.ustc.edu.cn, ymwang@ustc.edu.cn}

\begin{abstract}

In this paper, we analyzed a failed and a successful eruption that initiated from the same polarity inversion line 
within NOAA AR 11387 on December 25, 2011. They both started from a reconnection between 
sheared arcades, having distinct pre-eruption conditions and eruption details: 
before the failed one, the magnetic fields of the core region had a weaker non-potentiality; the external 
fields had a similar critical height for torus instability, 
a similar local torus-stable region, 
but a larger magnetic flux ratio (of low corona and near-surface region) 
as compared to the successful one. 
During the failed eruption, {a smaller Lorentz force impulse was exerted on the outward ejecta}; the ejecta had a much slower rising speed. Factors that might lead to the initiation of the failed eruption are identified: 
1) a weaker non-potentiality of the core region, and a smaller Lorentz force impulse 
gave the ejecta a small 
momentum; 2) the large flux ratio, and the local torus-stable region in the corona provided strong confinements that made the erupting structure regain an equilibrium state. 

\end{abstract}

\section{INTRODUCTION}\label{sec:intro}

Solar flares and coronal mass ejections (CME), as the most energetic phenomena in the solar atmosphere, 
are suggested to be different manifestations of a same magnetic explosive process when associated \citep[e.g.,][]{Harrison_1995, Lin_Forbes_2000}. 
{Their} association rate generally increases with the intensities and the durations of the flares. For example, in a survey that covers flares during 1995 - 2005, 
an association rate of $40\%$, $60\%$, $89\%$ in {\it GOES} C5.7-, M3.2-, X3-class flares, 
of $100\%$ in flares {of a duration} 
$>180$ minutes can be found \citep{Yashiro_etal_2006}. {However, }
exceptions like successive X-class flares without accompanied CMEs 
do exist \citep[e.g.,][]{Wang_Zhang_2007, Thalmann_etal_2015, Sunxd_2015, Lliu_2016}. 

Flares {accompanied by} 
CMEs are termed as ``eruptive" flares, while {those }
without 
are termed as ``confined" flares. 
Some confined flares {which} 
show a sign of ejecta that failed to fully erupt, are called ``failed/confined eruptions"\citep[e.g.,][]{Jihs_2003, Alexander_2006}. It should be noted that sometimes the observable ejecta may even appear in the inner region of the field of view (FOV) of coronagraphs, {being a ``CME" literally, but failed to propagate to a large distance in the corona.}
This kind of eruptions may not be magnetically driven, or may at least be different from the flux rope-related CMEs, therefore they are also defined as failed eruptions \citep{Vourlidas_2010, Vourlidas_2012}. 

{Successful }
eruptions, i.e.,  CMEs propagating into interplanetary space, may cause strong geomagnetic disturbances. {Hence, it is important for space weather forecasting to successfully identify the difference between explosive phenomena with and without successfully escaped ejecta. 
Extensive research on the subject of ``confined" and ``eruptive" flares, as well ``failed/confined" and ``successful" eruptions, has been done by e.g.,~\cite{Torok_Kliem_2005, Wang_Zhang_2007, Chengx_2011a, Sunxd_2015}.} 

{In most models, an eruption is usually driven by a core structure in form of a magnetic flux rope }
\citep[e.g.,][]{Tamari_etal_1999, Amari_2000, Roussev_2003, Torok_Kliem_2005}, {irrespective of }
the formation time (prior to or during the eruption) of the flux rope. The core structure can be tracked if some coupled plasma structures, e.g., sigmoid, filament, prominence, and hot channel, 
are observable \citep[e.g.,][]{Canfield_etal_1999, Schmieder_2013, Zhangj_2012}. Both theoretical and observational work confirm {that} helical kink instability {is} 
one of the onset mechanisms {for} 
eruptions, that will occur when the twist of the magnetic field lines exceed a critical value \citep[e.g.,][]{Torok_Kliem_2005, Guo_2010, Kumar_2012}; 
{torus instability is another important mechanism for full eruptions,} 
which will occur when the external fields decrease fast enough~\citep[e.g.,][]{Kliem_2006, Fany_2010}. 
Accordingly, {studies} 
on the magnetic conditions in active regions (ARs) mainly focus on two aspects. One is the non-potentiality of the core magnetic fields, such as twist number, electric current, magnetic free energy, and magnetic helicity, etc. \citep[e.g.,][]{Falconer_2002, Falconer_2006, Nindos_Andrews_2004, Sunxd_2015}. The other is the confinement of the external magnetic fields, measured by the decay of the magnetic fields with 
increasing height, i.e., decay index \citep{Torok_Kliem_2005}, or a {ratio of magnetic fluxes at different heights}~\citep{Wang_Zhang_2007}. 
Confined events usually have a weaker non-potentiality in the core region~\citep[e.g.,][]{Sunxd_2015, Lliu_2016}, 
and a stronger confinement in the low corona than successful eruptions \citep[e.g.,][]{Liuy_2008}. 
Besides, an eruption position close to the center of the AR \citep{Wang_Zhang_2007, Chengx_2011a, Chenhd_2015}, lack of opened or opening overlying magnetic fields~\citep{Jihs_2003}, {or} asymmetry of the magnetic backgrounds \citep{Liuy_2009}, may play roles in confining the eruptions.

The topology of the magnetic fields that {is} involved in the eruption, including that of the core fields and the overlying fields, will determine the eruption details. Therefore, a similar magnetic environment tends to produce similar eruptions. 
For example, homologous CMEs/flares, with similar eruptiveness, can occur from the same region within an AR, e.g., from the same polarity inversion line (PIL) \citep{Zhang_Wang_2002, Devore_2008, Chandra_2011, Wang_2013, Lliu_2017, Vemareddy_2017}. PILs are the boundaries of adjacent flux concentrations with inversed polarities. Core structures of the eruptions, i.e., flux ropes or sheared arcades, usually reside above the PILs. 
The same source PIL naturally hints a similar magnetic environment, but sometimes does not guarantee activities with similar eruptiveness. 
For example, \cite{Sheny_2011} studied a series of filament eruptions from the same source region, of which only one eruption successfully escaped the Sun. 
They found that the field strength at the low corona, decay index and asymmetry properties of the extrapolated overlying fields for the failed and successful eruptions had no significant difference, and argued that besides the confinement, the energy released in the low corona may also be crucial for an eruption to fully erupt. {Hence, further} 
comparative studies are needed.

In this paper, from the view of failed/successful eruptions, we present a comparative study between two eruptions that initiated from the same PIL within NOAA AR 11387. The first one was a failed eruption associated with a C8.4-class {\it GOES} soft X-ray (SXR) flare {(2011-12-25T11:20 UT)},  the second one was a successful eruption that escaped the Sun and evolved into a CME, associated with a M4.0-class flare {(2011-12-25T18:11 UT)}. 
We refer to the first, i.e., the failed (second, successful) one and its corresponding flare as eruption1 and flare1 (eruption2 and flare2) hereinafter. By performing a combined analysis, including eruption details and the evolution of the magnetic conditions based on stereoscopic observations, we try to discover the physical explanation for events that initiated from a possibly similar magnetic environment but with different eruptiveness. 
The paper is organized as follows: Data and methods are introduced in Section \ref{sec:data_meth}. Results are presented in Section \ref{sec:results}. Summary and discussion are given in Section \ref{sec:sum_discuss}.

\section{DATA AND METHODS}\label{sec:data_meth}

The two eruptions both occurred near {the coordinates} S20W20 
(see Table \ref{tb:paras} for details ), {and} are well observed by {\it SDO} \citep[{\it Solar Dynamics Observatory},][]{Pesnell_2012}. 
{\it STEREO-A }\citep[{\it Solar TErrestrial RElations Observatory-Ahead},][]{Kaiser_etal_2008} had a separation angle of $107^{\circ}$  from {\it SDO}, giving an additional limb-view for 
the eruptions. {We study the on-disk flaring evolution, using {\it SDO}/AIA~\citep[Atmospheric Imaging Assembly,][]{Lemen_etal_2012} data, that observes in seven EUV passbands and three UV passbands with a cadence up to 12 s and {and a resolution of} 0.6 arcsecs. 
We analyze the rising motion of the eruptions, using data from EUVI \citep[Extreme Ultraviolet Imager,][]{Wuelser_2004}, 
COR1 and COR2 \citep[Inner and Outer Coronagraphs,][]{Thompson_2003}, 
that are all comprised in SECCHI \citep[Sun Earth Connection Coronal and Heliospheric Investigation,][]{Howardr_2008} on board {\it STEREO-A}. 
We further apply a CME detection tool \citep{Bein_2011, Bein_2012} to track the eruption structure in EUVI and COR1, COR2 images,} by which its kinematics is obtained using a spline fit method.   

Besides the above data sets, photospheric vector magnetograms from {\it SDO}/HMI~\citep[Helioseismic and Magnetic Imager,][]{Hoeksema_etal_2014} 
are used to analyze the eruption-related magnetic conditions. {We take }
a subset of SHARPs \citep[Space-weather HMI Active Region Patches,][]{Bobra_2014} data products that automatically track an AR (or AR clusters). SHARPs are remapped from CCD coordinates into a heliographic coordinates with cylindrical equal area (CEA), with a resolution of 0.36~Mm per pixel and a time cadence of 720~s. 
Using SHARP data, we are able to calculate photospheric parameters, e.g., magnetic flux $\Phi$, shear angle $S$, current density $J_z$, current helicity $h_c$, energy density $\rho$, etc., in the whole AR or in a sub-region within the AR 
{(cf. the detailed calculation formulas in Table~\ref{tb:paras}). } 

Using DAVE4VM \citep[Differential Affine Velocity Estimator for Vector Magnetograms,][]{Schuck_2008} method to a time series of SHARP vector magnetograms, we can get the photospheric plasma velocity ($\bm{V}$). By subtracting the field-aligned plasma flow, $\displaystyle (\bm{V}\cdot{\bm{B}})\bm{B}/B^2$, from $\bm{V}$, the 
velocity 
perpendicular to the magnetic fields, {denoted by} $\bm{V}_{\perp}$, can be obtained.
Accordingly, the relative magnetic helicity flux through the photosphere can be calculated by: 
\begin{equation}\label{eq:helicity}
\frac{dH}{dt}\Bigg|_S=2\int_S(\bm{A}_p\cdot\bm{B}_t)V_{\perp n}dS-2\int_S(\bm{A}_p\cdot\bm{V}_{\perp t})B_ndS
\end{equation}
where $\bm{A}_p$ refers to the vector potential of the potential magnetic fields which have the same vertical component as the photospheric vector magnetic fields. $B_t$ ($B_n$) refers to the tangential (vertical) component of the fields, 
{and $V_{\perp t}$ ($V_{\perp n}$) refers to the tangential (vertical) component of $V_{\perp}$.} 
The left term of the above equation denotes the helicity injection rate contributed by the emergence of the twisted flux tube, while the right term refers to the helicity injection rate due to the shear motion on the photosphere \citep[e.g.,][]{Bergerm_1984, Liu_Schuck_2012, Liu_etal_2014}.

No direct observation of coronal magnetic fields is available. Thus, the three dimensional (3-D) coronal magnetic fields are reconstructed by 
a nonlinear force free fields (NLFFF) \citep{Wiegelmann_2004, Wiegelmann_2006, Wiegelmann_2012} and a potential fields (PF) extrapolation method~\citep[e.g.,][]{Sakurai_1989}, 
using the photospheric vector magnetograms as boundaries. Magnetic free energy can therefore be calculated by subtracting the magnetic potential energy from the total magnetic energy, i.e., with 
\begin{equation}\label{eq:ener}
E_F=E_{N}-E_P=\int_V\,\frac{B^2_{N}}{8\pi}dV-\int_V\,\frac{B^2_P}{8\pi}dV
 \end{equation}
in which $B_{N}$ refers to the NLFFFs and $B_P$ refers to the PFs, $dV$ denotes the elementary volume. The decay index that measures the decrease of the external magnetic fields with increasing height can be calculated by 
\begin{equation}\label{eq:decay}
n=-\frac{\rm \partial\,\ln B_{ex}(h)}{\rm \partial\,\ln h}
\end{equation} 
where $h$ refers to the height, $B_{ex}$ is the horizontal component of the external potential magnetic fields, since PFs play the major role in confining the eruption in torus instability theory \citep{Torok_Kliem_2005}. 
A ratio of the unsigned magnetic fluxes in {the} low corona ($h \approx 42$~Mm) and in the near-surface region ($h \approx 2$~Mm), which additionally measures the confinement of the overlying fields to the core region~\citep{Wang_Zhang_2007,Sunxd_2015}, can be calculated by $\Phi\,(42)/\Phi\,(2)$. 
$42$~Mm is found to be a typical height for eruption onset~\citep{Liuy_2008}. $\Phi\,(h)$ is calculated by $\sum |B_z\,(h)| dA$ in a plane with a height of $h$~Mm, {where} 
$B_z$ refers to the vertical component of the magnetic fields 
and dA 
to the elementary area of the plane.  

To study the magnetic parameters in the region that {is} most largely involved in the flare, a ``flaring PIL" (FPIL) is obtained 
following a method described in \cite{Sunxd_2015}: the PIL pixels are firstly located in a $B_z$ map, 
and then 
dilated with a {circular kernel ($r \approx 3$~Mm)}; the flaring pixels are firstly located in an AIA 1600\AA~ map near the peak time 
of the flare, with a threshold of $A_{m}+3 \times A_{d}$, where $A_m$ and $A_{d}$ are the mean value and the standard deviation of the 1600~\AA~image, 
{and} then 
dilated with the {kernel}; finally, the FPIL is determined as the intersection between the dilated PIL pixels and {the} flaring pixels. Within the region, the parameters in Table~\ref{tb:paras} are calculated. Besides, { 
the change of the vertical Lorentz force during the eruption, whose temporal integral can represent the force impulse that provides the outward ejecta's momentum, 
is calculated by: 
\begin{equation}\label{eq:force}
\delta F_{z,upward}=\frac{1}{8\pi}\int\,dA(\delta B_h^2-\delta B_z^2)
\end{equation}}
where $B_z$ ($B_h$) refers to the vertical (horizontal) component of the magnetic fields, and $dA$ refers to the elementary area 
\citep{Fisher_2012, Petrie_2012, Petrie_2013, Wanghm_2015}. 

With the data and method introduced above, we perform the analyzes. 

\section{RESULTS}\label{sec:results}

\begin{figure*}
\begin{center}
\includegraphics[width=0.65\hsize]{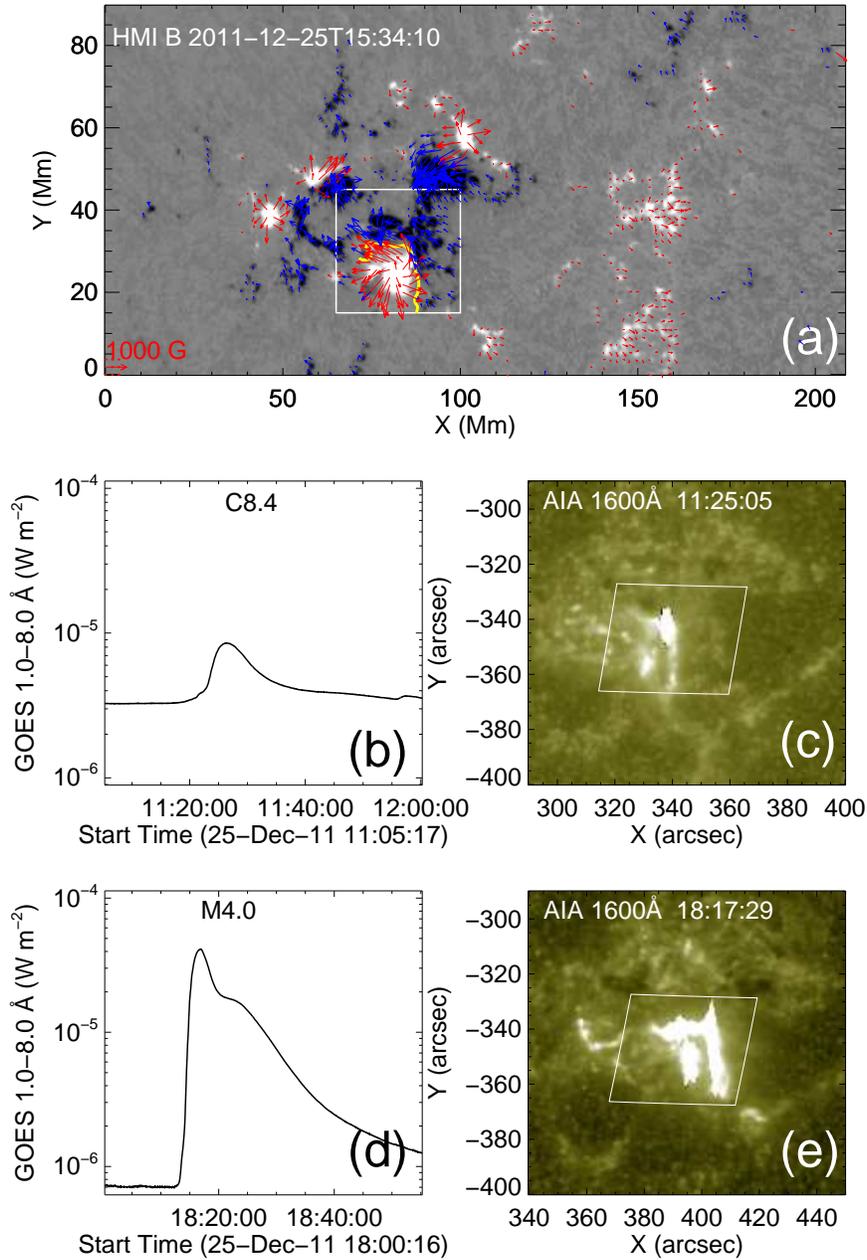}
\caption{Overview of the magnetic source and the associated flares. 
(a) vector magnetic fields of NOAA AR 11387 at a time instance between the two eruptions in the CEA heliographic coordinates; vertical component of the vector magentic fields ($B_z$) are plotted as background, white/black patches refer to the positive/negative $B_z$  that saturate at $\pm 800$~Gausses; red/blue arrows refer to the horizontial component of the vector magnetic fields ($B_h$) that originate from the positive/negative polarities; yellow line outlines the PIL that the two eruptions originated from; white box encloses the flux concentrations that form the PIL. (b), (d) are GOES SXR curves of the two flares, (b) for flare1 and (d) for flare2. (c), (e) are the corresponding observation in AIA/1600\AA~ near the flare peaks, showing the flare ribbons along the PIL, (c) for flare 1 and (e) for flare2. The white rectangulars in (c), (e) are the white box in (a) that remapped to the CCD coordinates.}  \label{fig:flr_b}
\end{center}
\end{figure*}

Figure~\ref{fig:flr_b} displays the magnetic source region and the accompanied flares of the two eruptions. 
The source AR had a multipolar configuration (as seen in Figure~\ref{fig:flr_b} (a)). Both eruptions were initiated from the same PIL (yellow line in Figure~\ref{fig:flr_b}~(a)), which was formed by two closely placed flux concentrations with opposite polarities (enclosed in white boxes in Figure~\ref{fig:flr_b}~(a),(c),(e)). The two flares had different magnitude, C8.4-class for flare1 and M4.0-class for flare2, {respectively} (Figure~\ref{fig:flr_b}~(b),(d)), but almost same durations (9 and 11 minutes). 
See more flaring details in Table~\ref{tb:paras}. They all produced flare ribbons along the source PIL (as seen in Figure~\ref{fig:flr_b} (c),(e)). 
For each eruption, we analyze their eruption details, pre-eruption magnetic conditions and eruption-related changes. 

\subsection{Eruption Details}

\begin{figure*}
\begin{center}
\includegraphics[width=0.9\hsize]{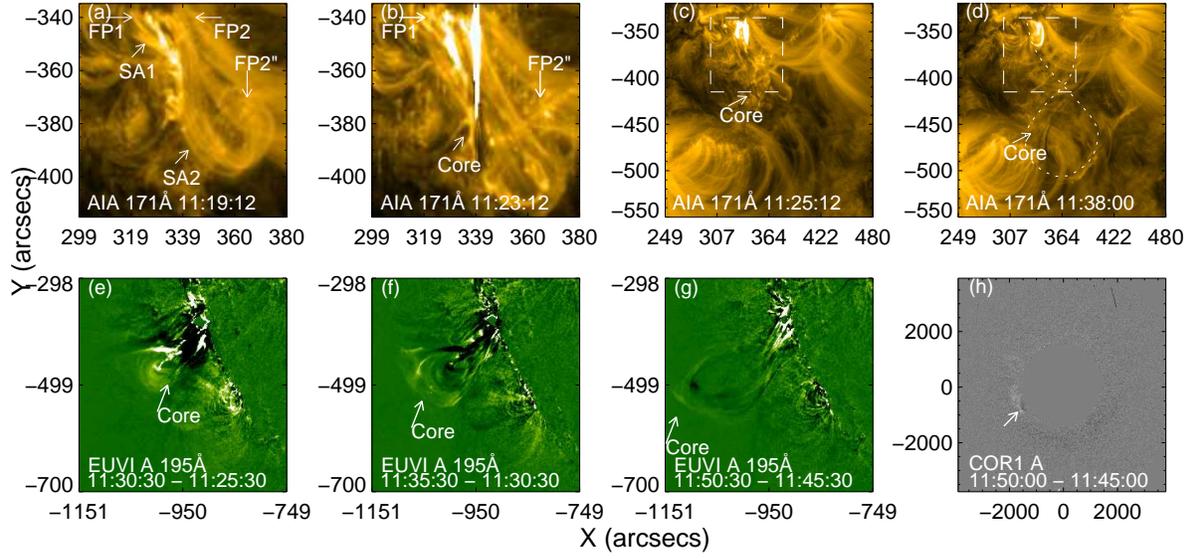}
\caption{Snapshots for the failed eruption. 
(a)-(d) flaring and eruption observed by AIA/171\AA. (e)-(g) eruption of the core structure observed by EUVI-A 195\AA~ from a limb view. (h) faint coronal outflow 
observed in COR1 A. Arrows SA1 and SA2 in (a) 
mark the two sets of sheared arcades involved in the eruption. Their footpoints are marked as FP1 (for SA1, short for footpoint1), FP2 and FP2$''$ (for SA2) in (a) and (b), respectively. 
Arrows named ``Core'' in (b)-(g) refer to the 
structure 
formed during the flare that drove the eruption but failed to erupt out. White dotted line in (d) outlines the $\gamma$ shape of the core structure during the eruption. 
Panels (c) and (d) have the same on-disk FOV, which is different from the FOV of (a) and (b). FOV of (a) and (b) is outlined by the white boxes in (c) and (d) for comparison. { An online animation lasts from 11:05 to 12:15, covering the eruption details 
not only shown above but also observed in AIA/131, 94, 304, 1600\AA, plus the GOES soft X-ray curve and evolution of the radial component ($B_r$) of the photospheric magnetic fields observed by HMI, is available. 
}}\label{fig:flr_cme_cas2} 
\end{center}
\end{figure*}

Figure~\ref{fig:flr_cme_cas2} shows the eruption process of the failed eruption (eruption1) observed by AIA (panels (a)-(d)), EUVI A (panels (e)-(g)) and COR1 A (panel (h)). Before the onset of the flare, two sets of sheared arcades (marked by arrows SA1 and SA2 in Figure~\ref{fig:flr_cme_cas2}(a)) were discernible in multiple wavelengths. The two sets of sheared arcades both lay above the source PIL. See more supporting evidence from the extrapolated magnetic fields in Section~\ref{subsec:flare-change}. 
The locations of the northern footpoints of the two sets of arcades (marked by FP1, FP2 in Figure~\ref{fig:flr_cme_cas2}, respectively) were very close.  Flaring between 
the two sets of sheared arcades can be observed. After the onset of the flare, mixture and exchange between the northern footpoints of SA1 and SA2 can be identified (see Figure~\ref{fig:flr_cme_cas2} (b)). A structure connecting the northern footpoint of SA1 (FP1 in Figure~\ref{fig:flr_cme_cas2} (a),(b)) and the {southern} footpoint of SA2 (FP2$''$ in Figure~\ref{fig:flr_cme_cas2}(a),(b)) formed and acted as the core structure of the eruption (see Figure~\ref{fig:flr_cme_cas2} (c),(d)). The structure showed a clear writhing motion,  which converted the twist of the plasma coupled field lines into the writhe of the axis of the flux-rope like core structure, 
then grew into a $\gamma$ shape in AIA observation (outlined by the white dotted line in Figure~\ref{fig:flr_cme_cas2}(d)), faded gradually, {and} finally became invisible. Accompanied plasma drainage {downward} to the solar disk can be observed. {For more details see the online animation}. The core structure showed {an} upward motion as observed by EUVI A with a limb view (Figure~\ref{fig:flr_cme_cas2} (e)-(g)) and finally halted at around 0.24 $R_{sun}$, and became invisible gradually. Meanwhile, extremely faint outflow can be observed in the inner boundary of the FOV of COR1 A (Figure~\ref{fig:flr_cme_cas2}(h)), which decelerated and diffused soon following the motion of the failed erupted 
core structure. {Concluding, this} 
can not be defined as a successful CME as introduced in Section~\ref{sec:intro}. 

The kinematical evolution of the core structure is shown in Figure~\ref{fig:fig_kinematics} (red curve), 
{and} is obtained by the CME detection tool introduced in Section~\ref{sec:data_meth}. 
The core structure was firstly observed by EUVI A 195\AA~at a height of $\sim$ 0.01 $R_{sun}$ above the solar surface, then slowly rose and stopped at the highest position around 0.24 $R_{sun}$. Accordingly, the velocity of the core structure peaked at the core's first appearance, then decreased gradually to 0. 
The inner corona outflow {that} appeared in COR1 A was extremely faint, diffuse without coherent shape or clear front, leading to a large uncertainty of its detection. 
Thus, its kinematics is not obtained. 

In summary, the failed eruption was driven by a flux rope-like structure that may form during the flare. The core structure rose slowly and finally halted/disappeared at around 0.24 $R_{sun}$.

Figure~\ref{fig:flr_cme_cas4} shows the eruption process of the successful eruption (eruption2) observed by AIA (panels (a)-(d)), EUVI A (panels (e)-(f)), COR1 A (panel (g)) and COR2 A (panel (h)). Before the onset of the flare, a sigmoid structure can be observed in multiple wavelengths of AIA, which is shown in 94~\AA~here (Figure~\ref{fig:flr_cme_cas4} (a)). Two sets of sheared arcades (marked by SA1 and SA2 in Figure~\ref{fig:flr_cme_cas4} (a)) can be identified. See more supporting evidence from the extrapolated magnetic fields in Section~\ref{subsec:flare-change}. The corresponding footpoints of the two sets of the sheared arcades (FP1 and FP1$''$ for SA1, FP2 and FP2$''$ for SA2 in Figure~\ref{fig:flr_cme_cas4} (b)) brightened after the flare onset. The intensity of the flaring kept increasing (Figure~\ref{fig:flr_cme_cas4} (c)), the sheared arcades or any structure {that} formed during the eruption can not be identified later. After the eruption, post-eruption loops can be observed (Figure~\ref{fig:flr_cme_cas4} (d)). {See more details in the online animation.}  
 
The kinematics of the CME in eruption2 is also checked and shown in Figure~\ref{fig:fig_kinematics} (in blue). 
The CME was firstly captured by EUVI A at a height around 0.1 $R_{sun}$, with a peak velocity of {1041} km/s, and then expanded {fastly} and entered the FOV of COR1 A before the flare ended. The velocity of the CME decreased as it propagated, and finally kept at a constant speed around 500 km/s.

In summary, the successful eruption {released} a CME rapidly. There may also be a reconnection between the two sets of sheared arcades during the eruption, similar as the process in the failed eruption, that needs further study based on coronal {magnetic} fields reconstruction (see {Section~\ref{subsec:flare-change}}).

\subsection{Pre-eruption Condition}\label{subsec:pre_eru}
\subsubsection*{Long-Duration Evolution of the AR}

In this section, we check the evolution of the source AR, NOAA AR 11387 in six hours before each eruption. Figure~\ref{fig:bxyz_velo} displays the snapshots of the evolving photospheric vector magnetic fields (Figure~\ref{fig:bxyz_velo} (a)-(c) for eruption1, (e)-(g) for eruption2) and six hours-averaged velocities $V_{\perp}$ 
(Figure~\ref{fig:bxyz_velo} (d) for eruption1, (h) for eruption2). Before both eruptions, clear magnetic flux emergence can be observed in the $B_z$ magnetograms. For eruption1, 
the AR already had a multipolar configuration six hours before its onset (Figure~\ref{fig:bxyz_velo}(a)). 
The source polarity pairs are highlighted in Figure~\ref{fig:bxyz_velo}~(white boxes and top right insets.) Within the region, the positive polarity (white patch) appeared and grew {fastly} during the six hours, accompanied by a westward movement that can be distinguished from its displacement. Meanwhile, the negative polarity (black patch) underwent a significant {morphological} change. The polarity pair {was} 
approaching each other. 
{The} $B_h$ component of the magnetic fields (red and blue arrows in Figure~\ref{fig:bxyz_velo} (a),(b),(c)) grew with the emergence of the polarities.  The {tangential component ($V_{\perp t}$}, orange arrows in Figure~\ref{fig:bxyz_velo}(d)) of the six hours-averaged  
velocities clearly {indicate} westward motion, while the  vertical velocities ($V_{\perp n}$, 
cyan contours in Figure~\ref{fig:bxyz_velo}(d)) {indicate} significant upflow at the east boundary of the positive polarity. The positive polarity moved westward faster than the negative one, indicating a net converging motion toward the PIL (yellow lines in Figure~\ref{fig:bxyz_velo} (c) and (d)) {(see online animation).} 

Before eruption2, flux emergence, especially for the positive flux, still existed in the source {polarity pairs} (Figure~\ref{fig:bxyz_velo}(e)-(g)). {Shortly} before eruption2 (Figure~\ref{fig:bxyz_velo}(g)), the polarities were much closer to each other 
than the ones at the time instance before eruption1 (Figure~\ref{fig:bxyz_velo}(c)); and $B_h$ 
ran more parallel to the PIL. {The} six hours-averaged tangential velocities also {indicate} a westward movement 
(orange arrows in Figure~\ref{fig:bxyz_velo}(h)), while the vertical velocities (cyan contours in Figure~\ref{fig:bxyz_velo}(h)) indicate a large extent upflow at the east boundary of the positive polarity, 
which was smaller compared to the upflow before eruption1 (Figure~\ref{fig:bxyz_velo}(d)). The positive polarity had a faster westward movement {and} 
brought the two polarities closer. {See online animation} for the evolution of the photospheric vector magnetic fields of the AR.  

The temporal evolution of some key parameters that characterize the properties of the entire AR are displayed in Figure~\ref{fig:para_evo}. 
The total, positive, negative magnetic flux (black, orange, cyan curves in Figure~\ref{fig:para_evo} (a)) increased with time, confirming the flux emergence as observed in the {magnetogram} snapshots (Figure~\ref{fig:bxyz_velo}). {The} flux change rate (dotted black curve in Figure~\ref{fig:para_evo} (a)) evolved, showing a larger value around $5\times10^{20}$~Mx/h before flare1  
than the value around $4.2\times10^{20}$~Mx/h before flare2. 
{The} helicity flux, i.e., {the} helicity injection rate calculated from {Equation}~\ref{eq:helicity}, including the total, emergence term and shear term are shown in Figure~\ref{fig:para_evo}~(b). 
The shear term (cyan curve in Figure~\ref{fig:para_evo}(b)) played a dominant role for the helicity injection.  
Before eruption1, the shear term had a mean value of $0.69\pm0.11\times10^{37}$~Mx$^2$/s, while the emergence term had a mean value of $0.43\pm0.07\times10^{37}$~Mx$^2$/s. 
Before eruption2, the shear term 
had a sudden increase after 14:00 {UT} but decreased again after 17:00 {UT}, with a peak value reaching $1.48\times10^{37}$~Mx$^2$/s. Meanwhile, the emergence term had a mean value of $0.50\pm0.12\times10^{37}$~Mx$^2$/s before 16:00 {UT}, comparable to the mean value before eruption1, and then showed a trend of decrease. Free magnetic energy is also calculated with {Equation}~\ref{eq:ener} based on the reconstructed coronal magnetic fields, shown in Figure~\ref{fig:para_evo}(c). {The magnetic} 
free energy increased before both {eruptions}. 
{More} free energy was accumulated {during the} 
six hours before eruption2 ($2.35\times10^{31}$~erg)  
than before eruption1 ($1.40\times10^{31}$~erg).

In summary, flux emergence, helicity injection and free energy accumulation existed in the AR before both eruptions. The {latter} 
two accumulated more remarkably before the successful eruption. 

\subsubsection*{Magnetic Conditions Prior to the Eruption Onsets} 

The magnetograms have a time cadence of 12 minutes. To explore the static, pre-eruption magnetic conditions, we choose the ones nearest and prior to the flare onsets, resulting {a} magnetogram with 10 minutes before flare1, and the other 
with 1 minute before flare2. 
On the photosphere, an FPIL mask (outlined by dotted green lines in Figure~\ref{fig:ener_dec} (a),(c)), a region that involves the source PIL and the flaring area, is obtained based on the combination of the magnetograms and the images in AIA/1600\AA~ for each eruption (see method details in Section~\ref{sec:data_meth}). We refer FPIL1 (FPIL2) as the mask for eruption1 (eruption2). Parameters in the FPIL masks are calculated and shown in Table~\ref{tb:paras}. In {general}, FPIL2 had a larger size than FPIL1, with an unsigned magnetic flux ($\Phi$) of $8.17\times10^{20}$~Mx, which was three times larger than for {FPIL1} ($2.71\times10^{20}$~Mx). 
{The} mean value of the strength of the magnetic fields ($\overline{B}$) in FPIL2 (814~G), which should not be affected by the mask size, was also larger than in FPIL1 (479~G). The parameters that measure the non-potentiality of the region all had larger values in FPIL2 than in FPIL1. See details in Table~\ref{tb:paras} (covering also parameters that measure the confinement of the external potential fields). 
The two events had similar critical heights (the heights where $n$ reach the {critical value, $1.5$, for torus instability}) below 15~Mm. 
{The ratio of magnetic fluxes between 42~Mm and 2~Mm for eruption2 (0.04) was lower than for eruption1 (0.08).} The free energy for eruption2 in the total computing 
volume ($4.71\times10^{31}$~erg) was larger than for eruption1 ($2.27\times10^{31}$~erg). 

For clarification, Figure~\ref{fig:ener_dec} displays the pre-eruption free energy map (Figure~\ref{fig:ener_dec}(a),(c)) of the core region {and the distribution of $B_h$ and decay index} 
above the FPIL (Figure~\ref{fig:ener_dec}(b),(d)). The free energy is integrated from the photosphere to 42~Mm. For both {eruptions}, the free energy had clear concentrations that covered the majority of the FPIL.  
The intensity, and the size of the free energy concentration for eruption2 {was} larger than 
for eruption1. Meanwhile, the distributions of the decay index above the FPIL ({orange} 
curves in Figure~\ref{fig:ener_dec}(b),(d)) showed similar variation trend of a ``saddle-like'' profile: for eruption1, $n$ reached the critical value of 1.5 at around 13.1~Mm, kept increasing and peaked at around 27~Mm with a value of 2, then decreased and dropped in a local torus-stable region within $[40,76]$~Mm, afterwards kept increasing with values larger than 1.5; 
for eruption2, a similar variation trend was found, though with slightly lower heights. $n$ reached the critical height at around 10.2~Mm, peaked at 21~Mm, and {fell} into the local torus-stable region within heights of $[36,67]$~Mm. 
For the median value of $B_h$ above the FPIL (black curves in Figure~\ref{fig:ener_dec}(b),(d)), the two distributions had a similar variation trend except larger values {near} the photosphere for eruption2. 

\begin{figure*}
\begin{center}
\includegraphics[width=0.85\hsize]{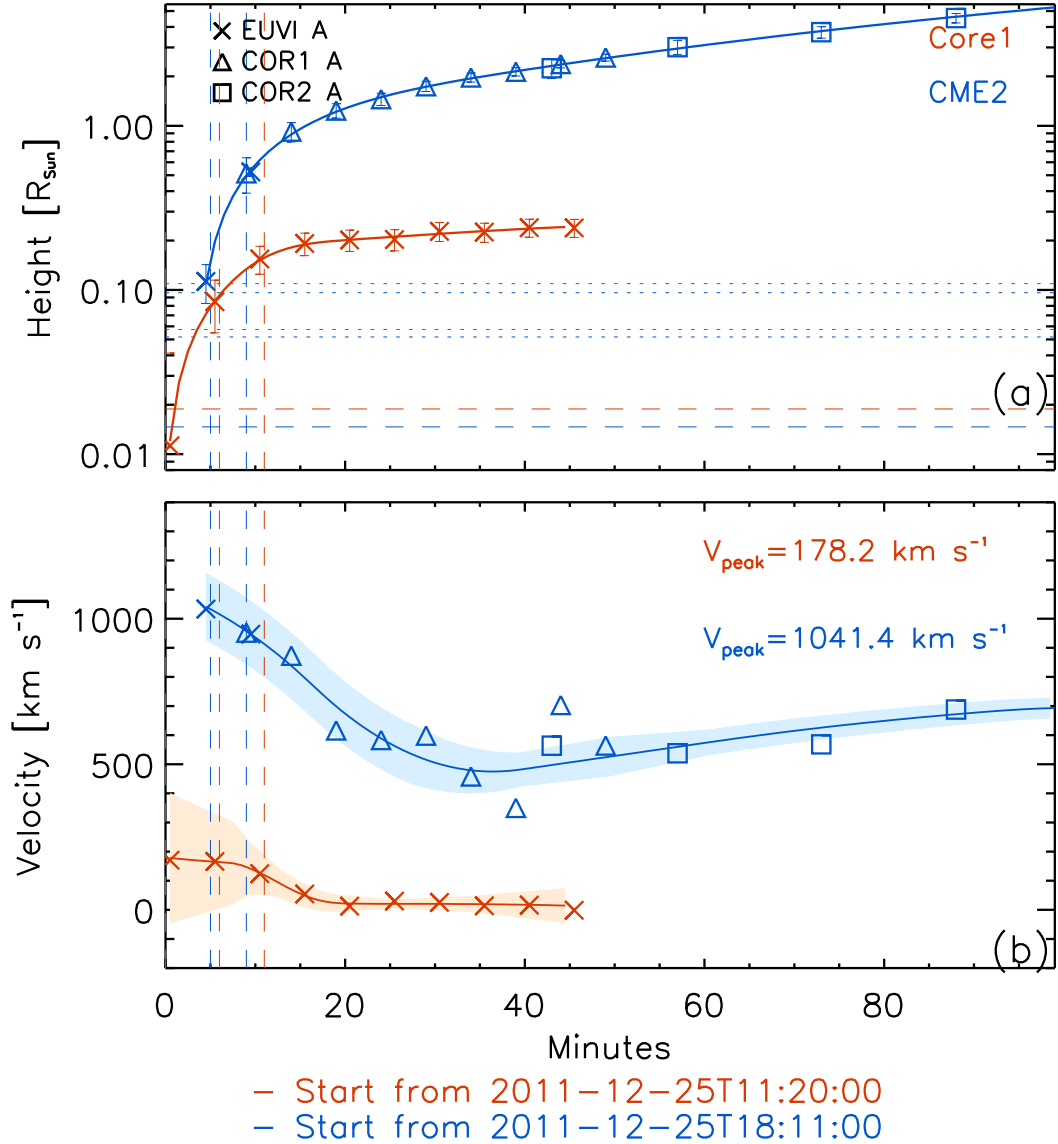}
\caption{Kinematics of the two eruptions. (a) heights of the eruptions. (b) velocities of the eruptions. The time axes all start from the instances of the flares onsets. The colored vertical dashed lines mark the peak and end times of the flares. The horizontal long dashed lines in (a) mark the critical heights, and the upper two sets of horizontal short dashed 
lines mark local torus-stable regions for each eruption (see details in Section~\ref{subsec:pre_eru}). The X marks are data points measured from {observations} of EUVI A, triangle symbols are for COR1 A and squares are for COR2 A. Height/velocity curves are fitted by a spline fit method {to} the observed data points. Red curves, points, texts give the information of the core structure of eruption1 (Core1);  
blue ones give the information of the CME in eruption2 (CME2). Uncertainties for the height measurments are overplotted as errorbars in (a), for the velocities are displayed as the colored, shaded regions in (b). }\label{fig:fig_kinematics} 
\end{center}
\end{figure*}

\begin{figure*}
\begin{center}
\includegraphics[width=0.85\hsize]{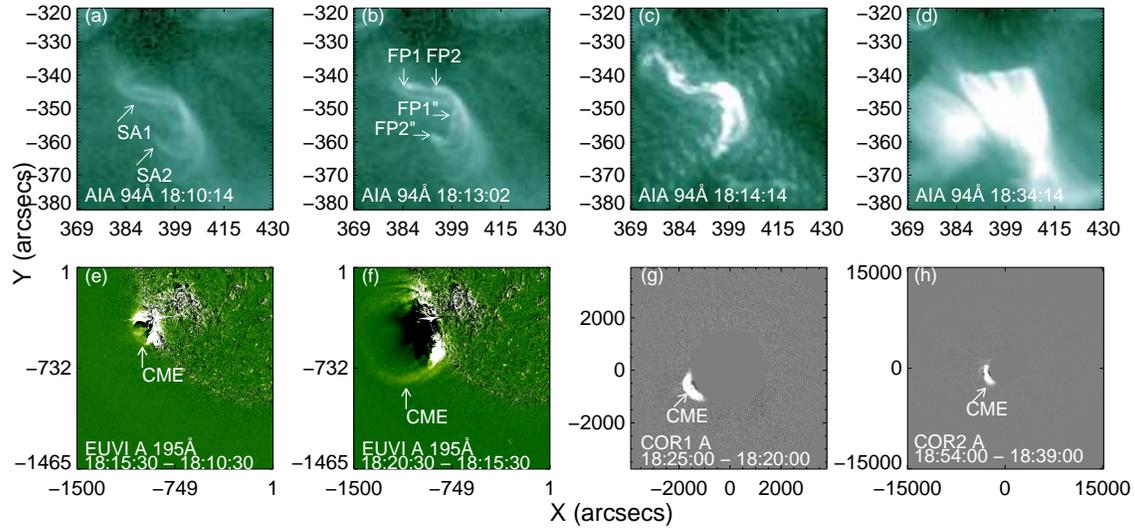}
\caption{Snapshots for the successful eruption. Similar layouts as Figure~\ref{fig:flr_cme_cas2}. (a)-(d) are the eruption process observed in AIA/94~\AA. Arrows SA1 and SA2 have similar meaning as the ones in Figure~\ref{fig:flr_cme_cas2}(a). Their corresponding footpoints are marked by arrows FP1, FP1$''$ and FP2, FP2$''$. The former pair are for SA1 and the later pair for SA2. 
(e)-(f): CME observed by EUVI A. (g)-(h): CME observed by COR1 A and COR2 A. { An online animation lasts from 18:05 to 18:50, showing the eruption details with similar layouts as the animation corresponding to Figure~\ref{fig:flr_cme_cas2}, is available. 
}}\label{fig:flr_cme_cas4} 
\end{center}
\end{figure*}

\begin{figure*}
\begin{center}
\includegraphics[width=0.75\hsize]{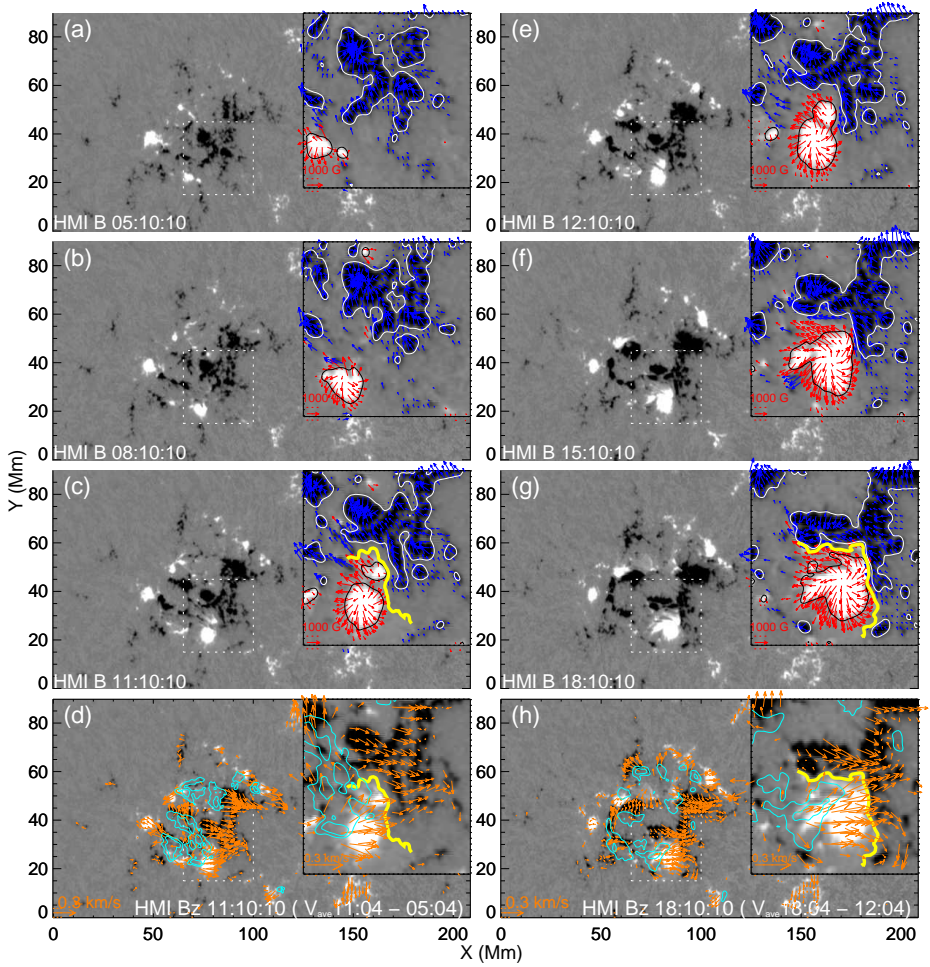}
\caption{Evolution of the photospheric vector magnetic fields, and the velocities, of the AR during six hours before each eruption. (a)-(d) 
evolution before eruption1. (e)-(f) 
evolution before eruption2. (a)-(c) maps of $B_z$ 
before eruption1 in every three hours. White dotted boxes have the same FOV as the white box in Figure~\ref{fig:flr_b}(a), enclosing the source magnetic polarity pairs, and are enlarged and displayed as insets on the right sides, with $B_h$ 
overplotted. The black (white) contours in the insets outline the {boundaries} of the positive (negative) polarity at $300$ ($-300$)~{G} for clarity. Colored background and arrows have the same meaing {as} the ones in Figure~\ref{fig:flr_b} (a). (d) shows the $B_z$ magnetogram (same as in panel (c)) {shortly} before the flare, with six hours-averaged $V_{\perp}$ 
calculated by DAVE4VM overplotted. The orange arrows display the horizontal component of $V_{\perp}$, while the cyan contours outline the vertical component of $V_{\perp}$ at $[0.05,0.09]$~km/s. The yellow lines in (c), (d) mark the PIL before eruption1.
(e)-(h) have the same layouts as (a)-(d) but for eruption2. { An online animation, showing the 12 hours evolution of the photospheric vector magnetic fields, is available. 
}}\label{fig:bxyz_velo}
\end{center}
\end{figure*}

\begin{figure*}
\begin{center}
\includegraphics[width=0.70\hsize]{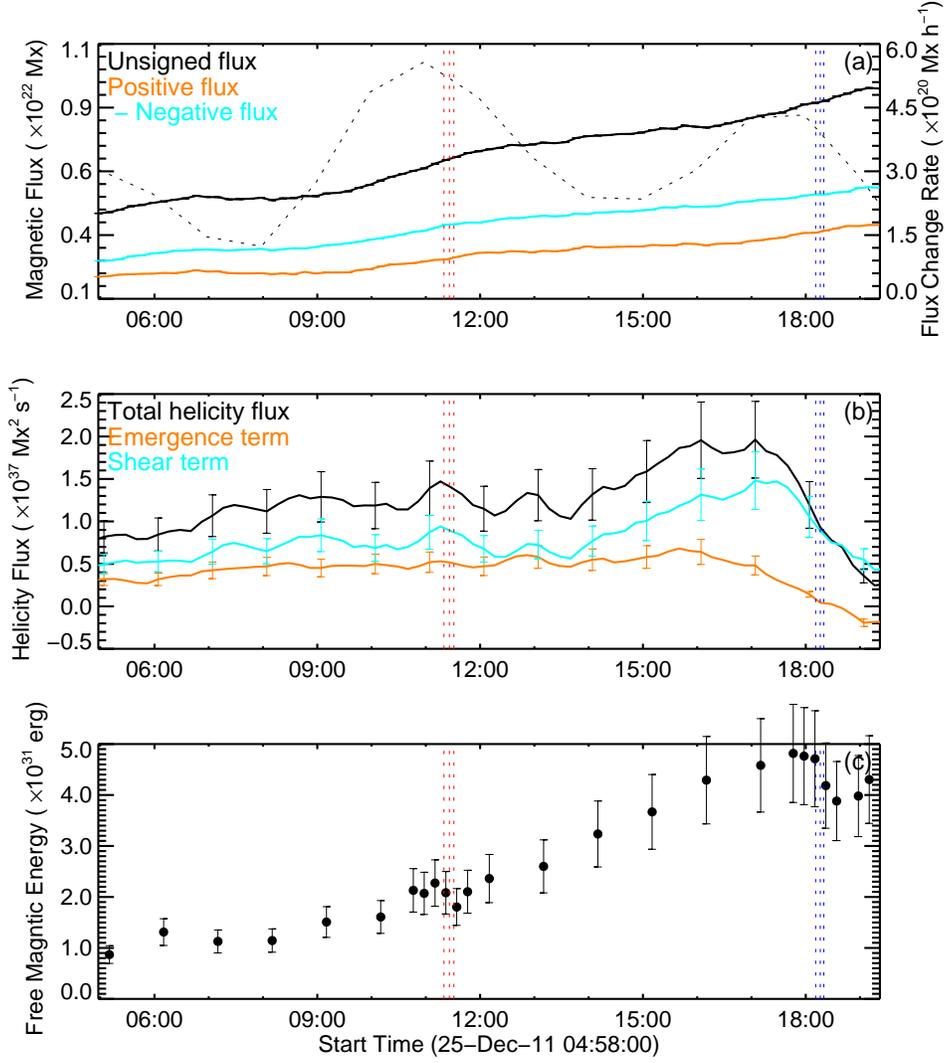}
\caption{Evolution of magnetic flux, helicity injection rate and the free magnetic energy of the AR, covering the durations of six hours before eruption1 and six hours before eruption2. (a) evolution of magnetic fluxes, black solid line for the total unsigned magnetic flux, black dotted line for the magnetic flux change rate, orange line for the positive magnetic flux and cyan line for the absolute value of the negative magnetic flux, which is suited for a direct comparison. The uncertainties of the magnetic flux calculated from the error data segment from HMI are overplotted on the curves, which are too small to be distinguished. (b) one hour-averaged helicity flux calculated by {Equation}~\ref{eq:helicity}, black line for the total helicity flux, orange line for the helicity flux from the emergence term and cyan line for the helicity flux from the shear term. An uncerntaity, of $23\%$ of the helicity flux value \citep[a value from][]{Liu_Schuck_2012} is overplotted every 5 data points. (c) free magnetic energy calculated in the {reconstructed} magnetic fields by {Equation}~\ref{eq:ener}. Uncertainties of $20\%$ of the free energy values are overplotted~\citep{Thalmann_2008a}. The energy data points have a cadence of 720s during the eruption, and 1 hour during other durations. In all panels, red dotted lines mark the beginning, peaking and ending time of the flare1. Blue dotted lines give same information but for flare2. }\label{fig:para_evo} 
\end{center}
\end{figure*}

\begin{figure*}
\begin{center}
\includegraphics[width=0.9\hsize]{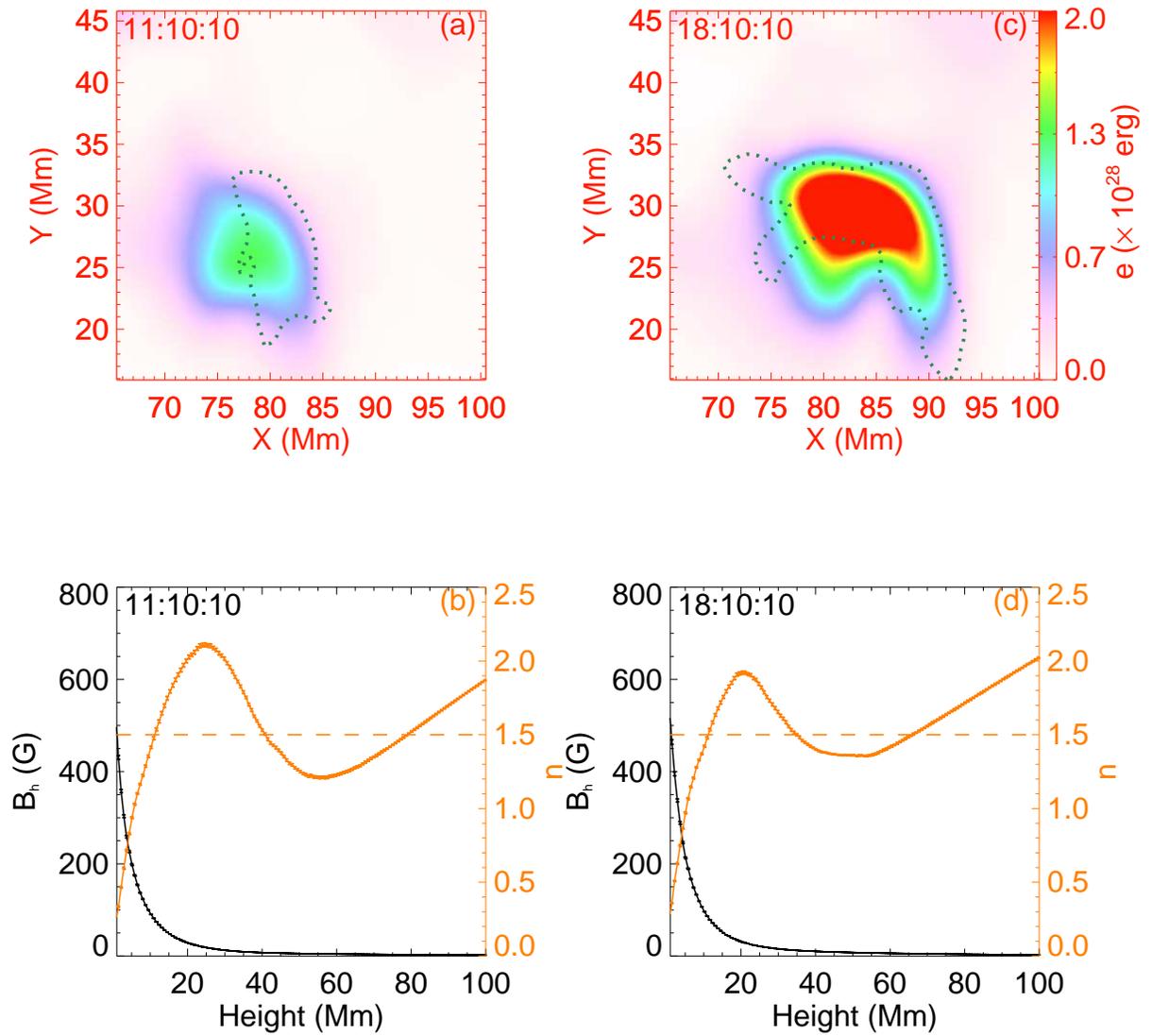}
\caption{Coronal magnetic conditions {shortly} before the two eruptions. (a),(b) for eruption1, (c),(d) for eruption2. (a) pre-eruption1 magnetic free energy map with the same FOV {as} Figure~\ref{fig:flr_b}(a). 
Each pixel represents the free energy integrated in a volume that takes 
the pixel at the photosphere, and the one at the height of 42~Mm as the lower and upper boundaries. The energy value saturated at $[0, 2.0\times10^{28}]$~erg. The dotted green lines outline the FPILs. (b) $B_h$ and decay index $n$ distribution from 0 to 100~Mm, are shown as the black and orange curves, {respectively}, with standard error overplotted. Values at each height are the median values in the FPIL mask region. The position where $n$ reach the critial value 1.5 is outlined by the orange dashed line. (c),(d) have the same layout as (a),(b) but for the time instance before eruption2.}\label{fig:ener_dec} 
\end{center}
\end{figure*}

For each eruption, the {pre-eruption} critical height (long dashed horizontal line in Figure~\ref{fig:fig_kinematics} (a)) and 
torus-stable region (enclosed in the upper two short dashed horizontal lines in Figure~\ref{fig:fig_kinematics} (a)) {are marked in Figure~\ref{fig:fig_kinematics} for comparison.}  
The core structure of the failed eruption had an initial height of $\sim$ 0.01 $R_{sun}$ ($ \approx 7$~Mm). It rose rapidly in the lower torus-unstable region, and slowed down to a large extent after it passed the torus-stable region. The CME in the successful eruption rose much faster than the failed one, and passed the torus-stable region more quickly. 

In summary, before the eruption onset, the core region of the successful eruption displayed larger non-potentiality than 
the failed one. Their decay index distributions had a similar variation trend like a ``saddle'',  in which a local torus-stable region was enclosed by two torus-unstable regions. {The torus-stable region may play a role in confining the failed eruption.}

\subsection{Eruption-related Change}~\label{subsec:flare-change}

\begin{figure*}
\begin{center}
\includegraphics[width=0.95\hsize]{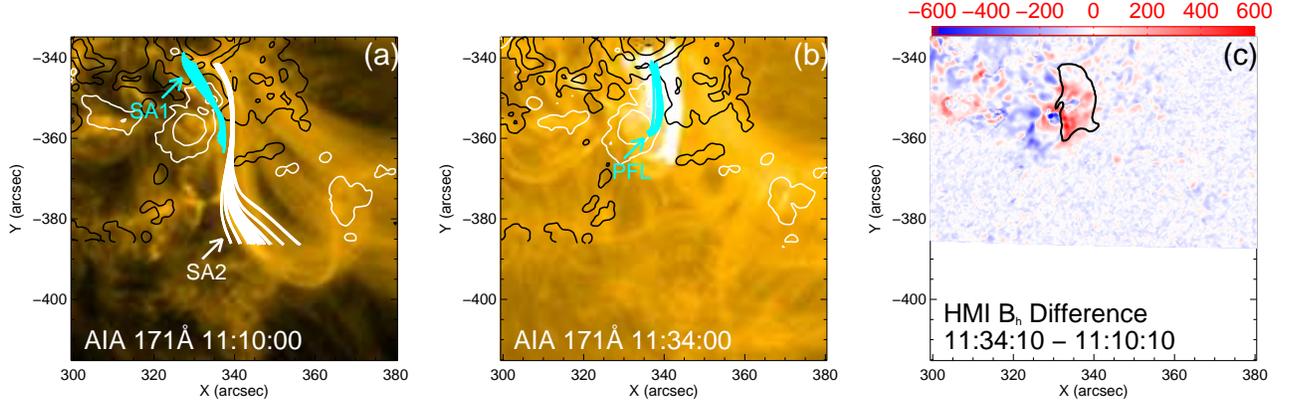}
\caption{Eruption1-related change. (a),(b) have the same FOV as in Figure~\ref{fig:flr_cme_cas2}(a)-(b). SA1 (sheared arcade1) and SA2 in (a) are two sets of sheared field lines extracted from the reconstructed coronal magnetic fields based on NLFFF method at the time instance before the eruption. The field {lines} are overplotted on the image taken by AIA 171\AA~ at the same time. PFL in (b) refers to the post-eruption loops extracted from the model coronal fields after the eruption. The background of (b) is the observation in AIA 171\AA~ at the same time.  White/black contours in (a) and (b) outline the photospheric $B_z$ 
at $[800,200,-200,-800]$~{G}. (c) {shows} the difference between the photospheric $B_h$ 
before and after the eruption, which saturates at [-600,600]~{G}. The black curve outlines the FPIL for eruption1. The magnetogram in (c) is projected from the CEA heliographic coordinates to the CCD coordinates, having the same FOV as (a) and (b).}\label{fig:nlfff_bh_cas2} 
\end{center}
\end{figure*}

A flux rope-locating method, using the combination of the twist number and the squashing factor Q calculated from the reconstructed 3-D coronal magnetic fields \citep{rliu_2016}, is performed to find the possible pre-existing flux rope. However, no coherent pre-existing flux rope can be located before both eruptions. {Therefore the} result is not presented here. 
We then checked the detailed configuration of the magnetic fields above the source PIL in the extrapolated coronal fields before and after the eruptions. For eruption1, the result is shown in Figure~\ref{fig:nlfff_bh_cas2} (panels (a),(b)). Before the eruption, two sets of sheared field lines (SA1 and SA2 in Figure~\ref{fig:nlfff_bh_cas2} (a)) that corresponded well with the sheared arcades observed by AIA (see background of Figure~\ref{fig:nlfff_bh_cas2} (a), or SA1 and SA2 in Figure~\ref{fig:flr_cme_cas2} (a)) were found. Note, the coronal magnetic fields are reconstructed using the HMI SHARP cutout magnetograms as photospheric boundaries, which {limits} 
the extrapolation cube {region so} 
that sometimes the extrapolated field lines may go in and out of the boundary, as the field lines of SA2 {in Figure~\ref{fig:nlfff_bh_cas2} (a)} showed: their southern part was out of the extrapolation box. However, the northern part {(left in the box)} coincided well with 
the corresponding arcades observed by AIA (SA2 in Figure~\ref{fig:flr_cme_cas2}(a)). The extrapolated fields also met the divergence-free and force-free condition required by the NLFFF method (see details in APPENDIX~\ref{subsec:quali_nlfff}). 
Thus, we still take the result. 
After eruption1, the two sets of the sheared field lines can not be identified, 
neither in the AIA observation (background of Figure~\ref{fig:nlfff_bh_cas2}(b)) nor in the reconstructed magnetic fields. On the contrary, near-potential loops can be identified in both, observation and the model corona (cyan field lines marked as PFL in Figure~\ref{fig:nlfff_bh_cas2}(b)). 
Combined with the evolution of the eruption, we argue that a 
reconnection between the two sets of the sheared {arcades} that formed the eruption core, may have happened. 

Besides checking the topology change, we also checked the change of the photospheric $B_h$ as shown in Figure~\ref{fig:nlfff_bh_cas2}(c). The map is projected into CCD coordinates to be compared with Figure~\ref{fig:nlfff_bh_cas2} (a), (b). A clear enhancement of $B_h$ was discernible in the FPIL region (outlined by a black curve in Figure~\ref{fig:nlfff_bh_cas2}(c)). 
The change of the {Lorentz force}, 
calculated by {Equation}~\ref{eq:force}, gave a value of $0.11\times10^{22}$ {dyn}. Meanwhile, the magnetic free energy showed a {decrease} of $0.47\times10^{31}$~erg (see Table~\ref{tb:paras}). 

\begin{figure*}
\begin{center}
\includegraphics[width=0.95\hsize]{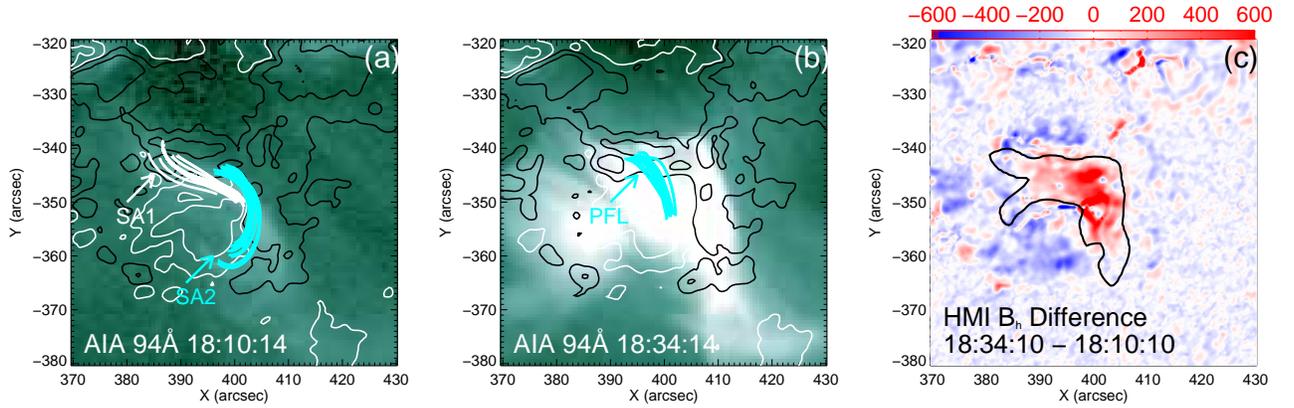}
\caption{Eruption2-related change. Similar layouts as Figure~\ref{fig:flr_cme_cas2}. Backgrounds of (a) and (b) are observations taken in AIA~94\AA.}\label{fig:nlfff_bh_cas4} 
\end{center}
\end{figure*}

For eruption2, similar changes are derived and shown in Figure~\ref{fig:nlfff_bh_cas4}. Before the eruption, two sets of sheared field lines (SA1 and SA2 in Figure~\ref{fig:nlfff_bh_cas4}~(a)) above the PIL can be identified, which corresponded well with the position of the sheared arcades observed by AIA (backgrounds in Figure~\ref{fig:nlfff_bh_cas4}~(a) and SA1, SA2 in Figure~\ref{fig:flr_cme_cas4} (a)). After eruption2, the two sets of the sheared field lines were not discernible anymore, and near-potential 
loops were found above the source PIL (exemplary loops are shown as PFL in Figure~\ref{fig:nlfff_bh_cas4}~(b)). Besides the topology change of the fields, the enhancement of the photospheric $B_h$ 
in the FPIL can also be recognized (Figure~\ref{fig:nlfff_bh_cas4}~(c)). 
The change of the {Lorentz force} had a value of $1.64\times10^{22}$ dyne, which was an order of magnitude larger than the value of the force change in eruption1. The magnetic free energy also showed a larger decrease of $0.83\times10^{31}$~erg (shown in Table~\ref{tb:paras}). 

In summary, {the derived} topology changes in the reconstructed magnetic fields support that for both eruptions, a reconnection between the sheared arcades above the source PIL may have happened. Compared to the failed eruption, the successful eruption revealed a larger change in the Lorentz force and a stronger decrease in the magnetic free energy.

\section{SUMMARY AND DISCUSSION}\label{sec:sum_discuss}
\subsection{Summary}

In this work, we perfomed a detailed comparative study between a failed and a successful eruption that initiated from the same PIL within NOAA AR 11387 from perspectives of their eruption details, pre-eruption magnetic conditions, and the eruption-related changes by stereoscopic observation from {\it SDO} and {\it STEREO-A}. The results are summarized as follows:

\begin{enumerate}
\item For the failed eruption (eruption1), two sets of sheared arcades above the source PIL can be identified from AIA {observations} before the eruption. The sheared arcades were likely to reconnect during the flare, 
and led to the formation of a flux rope-like core structure that drove the outward eruption. 
The structure rose slowly with writhing motion and mass drainage, and finally stopped, became {gradually invisible afterwards}. The process fits into a scenario of a failed eruption of a helical kinked flux rope \citep[e.g.,][]{Fan_2005, Torok_2010, Guo_2010, Hassanin_2016}. The core structure had a peak velocity of {178~km/s}, 
and ceased at a height around 0.24 $R_{sun}$. 

For the successful eruption (eruption2), two sets of sheared arcades above the PIL were also identified 
before its onset. Their corresponding footpoints {brightened} 
during the flare, 
indicating a reconnection between them. 
The eruption rapidly evolved into a fast CME (with a peak velocity of {1041 km/s}) that {propagated} into the heliosphere. 
\item  Before both eruptions, continuous flux emergence existed in the source AR. For the source polarity pair, the positive one displayed a faster westward motion 
than the negative one, resulting {in} a net convergence toward the PIL. 
{Due to} continuous flux emergence and shear motion on the photosphere, magnetic helicity was 
injected, 
with {a dominating} 
shear term. 
A larger quantity of magnetic free energy and magnetic helicity was accumulated before the successful eruption.  

Before the {onset of the} eruptions, 
the core region, i.e., the source FPIL of the successful eruption displayed a larger non-potentiality than the failed eruption. For example, {before the successful eruption,} more magnetic free energy resided above the FPIL, {and larger values were derived for} the mean current density, current helicity, and the shear angle that measures the core region's non-potentiality. 
{The} decay index distributions showed no significant difference, although the critical height for the successful erupion was slightly lower (around 3~Mm) than for the failed eruption. 
The ratio of magnetic fluxes at 42~Mm and 2~Mm (in the FPIL mask) was smaller before {the} successful eruption, indicating a relatively weaker confinement. 

\item Before the failed eruption, two sets of sheared field lines that corresponded well with the observed sheared arcades, can be identified in the coronal magnetic fields extrapolated by {the} NLFFF method. After the eruption, the sheared field lines disappeared, while near-potential loops were identified above the PIL. Combined with the observed eruption process, we conjecture that a reconnection between the sheared arcades may have {occurred}. {A similar} 
topology change was found during the successful eruption. 

Significant enhancements of $B_h$ were found in the FPILs after both eruptions, while the value of the Lorentz force change {during} 
the successful eruption was an order of magnitude larger than for 
the failed eruption. Decreases of magnetic free energy were also found after both eruptions, 
although the magnitude 
for the successful eruption was larger than {for} 
the failed one. 
\end{enumerate}

\subsection{Discussion}
The associated flares of the two eruptions had different intensities, C8.4-class and M4.0-class for the failed and {the} successful {one}. 
In general, the CME association rate increases when flare intensity increases, but the rate values for C8.4-class and M4.0-class are $45\%$ and $65\%$, respectively, \citep[see figure 1 in][]{Yashiro_etal_2006}, {which} are comparable. 
The event choice here is {therefore} adequate. 
During the failed eruption, an extremely faint corona outflow, without coherent shape or clear front, appeared in the FOV of COR1, soon diffused and failed to travel to a distance larger than 1 $R_{sun}$. Observation of COR2 and LASCO \citep{Brueckner_1995} confirmed the absence of a successful CME. This kind of literal ``coronal mass ejections'' that appeared in the inner corona but failed to propagate to a large distance, 
are defined as ``pseudo-'' or ``failed-'' CMEs in \cite{Vourlidas_2010, Vourlidas_2012}, which are thought to be different from the flux-rope related CMEs, {as they} 
may not be magnetically driven. Therefore, we argue that defining eruption1 as a failed eruption is reasonable. The physical nature of this kind of {``pseudo-'' CME is worth} a further study. 

Enhancement of $B_h$, which was observed during both eruptions, 
consists with the ``magnetic implosion'' scenario, in which the enhanced $B_h$ is thought to be attributed to the 
contraction of the field loops due to the decrease of the magnetic pressure resulted by the eruption \citep[e.g.,][]{Hudsons_2000, Wanghm_2010, Liuc_2012c, Wangs_2012b}. 
When no eruption happens, the solar atmosphere is in a quasi-static, i.e., roughly force-free state. The state does not stand during the eruption. Using the change of the magnetic fields during the eruption, the change of the upward Lorentz force { exerted on the 
ejecta can be calculated by {Equation}~\ref{eq:force}, of which the temporal integral will represent the Lorentz force impulse. 
}A larger impulse results in a larger momentum of the outward ejecta, indicating 
a faster velocity if the mass of the ejecta {is} 
comparable \citep[Equation 14 in][]{Fisher_2012}. { 
For the above two eruptions, {although precise times during which the Lorentz force acting on the ejecta can not be obtained, their associated flares have comparable durations, thus 
the Lorentz force impulses may 
be reflected directly by the changes of the force. 
}
}
During the successful eruption,
an order of magnitude larger Lorentz force change  
was found, consisted with the result of \cite{Sunxd_2015}, in which the Lorentz force change 
in an X3.1-class confined flare from NOAA AR 12192 was smaller than in 
other eruptive flares. 
{Furthermore}, the successful eruption had a larger peak velocity in the low corona than the failed eruption. We therefore argue that a 
{ 
large Lorentz force impulse, 
recorded by the Lorentz force change, may be important} 
for a successful eruption, which will give the medium weight ejecta a large initial velocity, make it quickly enter the region with weak confinement and escape the Sun. {  
The Lorentz force change, as a easily computed parameter, 
}has been confirmed by \cite{Wangs_2012a} to be linearly correlated to the magnitude of the flares. Its correlation to the eruptiveness of the flares is worth a further statistical study. 

{ It should be noted that the Lorentz force change (or force impulse) may not be a sufficiently independent parameter. Considering the original form of the equation for Lorentz force: $\bm{F}=\int_V \bm{J}\times \bm{B}\,dV$, a non-zero force firstly requires the existence of current in the volume, which is related to the non-potentiality of the magnetic fields; it secondly requires the current density to have a component perpendicular 
to the magnetic field lines, i.e., is additionally influenced by the fields configuration. Besides, the force change can only be obtained after the eruption, that may not be practical to pre-evaluate the potential of a source region for producing CMEs. However, 
it still reflects the property of the force acting on the ejecta during the eruption, 
that may also be 
important as the pre-eruption conditions of the source region 
in determining the final state of the ejecta.}

Before both eruptions, the decay index distribution displayed a ``saddle-like'' profile, exhibiting a local torus-stable ($n<1.5$) region higher than the critical height, enclosed by two torus-unstable domains. This kind of profile is found to be {exclusive} in ARs with multipolar configuration, and may provide extra confinement when the erupting core structure enters this torus-stable region without {a} well-developed disturbance~\citep{Wangd_2017}. In our failed eruption, the erupting core reached a height around 0.24 $R_{sun}$, which was higher than the critical height or the local torus-stable region. 
It rose quickly in the lower torus-unstable region, but slowed down largely after passing through the torus-stable region. 
{It} exhibited a writhing motion that converted the twist to the writhe of the structure axis, which {is} 
a typical behavior in the course of helical kink instability, suggesting a flux rope configuration of itself, 
and a possible ``self-consistent'' reformation. 
It may have regained an equilibrium state due to the stronger confinement in the torus-stable region, rose {more slowly} 
and finally halted. 
For the successful eruption, the erupting structure had larger energy/initial velocity to pass through the torus-stable region rapidly, {and was }
kept in a non-equilibrium state with enough disturbance to erupt out. The local torus-stable region in our case, which is also found in some other failed eruptions \citep[e.g.,][]{Guo_2010, Wangd_2017}, may play an important role in confining the eruption, especially those ones with small initial momentum. The result suggests that 
the role of the decay index in determining a full eruption {should be considered by} 
its entire distribution, rather than a single critical height. 

Before both eruptions, no pre-existing coherent flux rope can be found. 
However, sheared arcades were found above the PIL, which may have reconnected and initiated the eruptions. This is consistent with the result in ~\cite{Lliu_2016} who concluded that pre-existing flux ropes or sheared structures 
are necessary conditions for successful CMEs, although not sufficient ones. 

In summary, we analyzed two eruptions initiated from the same PIL, 
one with a failed erupted core and a faint inner corona outflow that {is} defined as a failed eruption, one with a fully evolved CME that {is} defined as a successful eruption. 
They both started from 
reconnection between different sets of sheared arcades above the source PIL due to converging motion, during which the flux ropes may have formed and driven the eruptions. The successful eruption had a larger velocity than the failed one. Although originated from the same PIL, they had distinct pre-eruption magnetic conditions: 
for the successful eruption, the source region underwent a more severe shear motion, accumulated more helicity and free energy, leading to a core region with a larger non-potentiality before the eruption started;   
the external magnetic fields displayed a similar decay index distribution (a ``saddle-like'' profile) but smaller flux ratio (between values in the planes at 42~Mm and 2~Mm) compared to the conditions for the failed eruption. 
{ The Lorentz force change exerted 
on the outward ejecta over the course of the eruption, which can represent 
the force impulse,} was an order of magnitude larger than for the failed eruption. 
We argue that {the} weaker non-potentiality in the core region, smaller {Lorentz force impulse} during the eruption, and the local torus-stable region in the coronal magnetic fields are together responsible for the failed eruption: the core structure 
erupted with a small momentum due to weaker non-potentiality {in its source and a small Lorentz force impulse exerted 
on it, }
may regain an equilibrium state 
due to the strong confinement in the torus-stable region, and thus failed to fully erupt. { The Lorentz force impulse during the eruption (which may be related to the non-potentiality and the configuration of the source fields),} and the local torus-stable regions in the corona may play important roles in initiating and confining the eruptions. 

\begin{table*}
\begin{center}
\caption{Characteristics of the two eruptions}\label{tb:paras}
\renewcommand\arraystretch{0.95}
\begin{tabular}{ccp{1.4cm}<{\centering}cccc}
\hline
 &  &  & Event1 & Event2   & Unit \\
\hline
Flare  & Begin & & 2011-12-25T11:20 UT & 2011-12-25T18:11 UT  & \\ 
       & Peak & & 11:26 & 18:16 & \\ 
       & End  & & 11:31 & 18:20 & \\
       & Duration & & 11  & 9 & min \\
       & Location & & S23W22 & S22W26 &  \\
\hline
Eruption$^a$ & & & Confined & Eruptive & \\
         & Core & $H_h$ & $0.24$   & - &  $R_{sun}$ \\
         &  & $V_{peak}$ & $178.2$   & - &  $km\ s^{-1}$ \\
         & CME & $H_h$ & - & $> 13 $ &  $R_{sun}$ \\
         &  & $V_{peak}$ & - &$1041.4$ &  $km\ s^{-1}$ \\
\hline
Parameters$^b$ & FPIL & $\Phi$ & $2.71 $ & $8.17 $  & $10^{20}Mx$ \\
           &      & Area & 92.75 & 185.76 & $Mm^{2}$ \\
           &      & $\overline{B}$ & $479\pm 13$ & $814\pm 10$  & G \\
           &      & $\overline{S}$ & $50.88 \pm 0.03$ & $56.80\pm 0.02$  & Degree \\
           &      & $\overline{J_z}$ & $0.61\pm0.02 $ & $-1.86\pm0.02 $  & $mA\ m^{-2}$ \\
           &      & $\overline{h_c}$ & $0.05\pm0.00 $ & $0.13\pm0.00 $  & $G^2 m^{-1}$ \\
           &      & $\overline{\rho}$ & $1.61\pm0.00 $ & $4.31\pm0.00 $  & $10^{19}erg\ cm^{-3}$ \\
           & Corona& Critical height &$13.11\pm0.73$  & $10.12\pm0.73$ & Mm  \\
           &  &  $\Phi(42)/\Phi(2)$ & $0.08$ & $0.04$ & \\
           &  & $E_f$ & $2.27\pm0.45$ & $4.71\pm0.94$ & $\times10^{31}$ erg \\
\hline   
Change  &      &$\Delta B_{h}$& +124 & +160 &G\\  
        &      &$\Delta E_{f}$& -0.47 & - 0.83 &$10^{31}erg$\\
        &      &$\Delta F_z$& +0.11 & +1.63 &$10^{22}dyn$\\ 
\hline
\end{tabular}
\end{center}
$^a$ $H_h$ represents the highest height the eruption structure reached, $V_{peak}$ represents the peak velocity of the eruption. ``-'' means no result is obtained. \\
$^b$ Unsigned magnetic flux $\Phi (h)$ is calculated by $\sum|B_z (h)|dA$ in the plane intersected by the FPIL mask, h refers to an abitrary height. 
 Mean shear angle $\overline{S}$ is computed from $ \frac{1}{N}\sum arccos(\frac{\bm{B}_o\cdot\bm{B}_p}{|B_o||B_p|})$, 
where $B_o$ ($B_p$) denotes the observed (potential) fields. Mean current density $J_z$ is calculated by $\frac{1}{N}\sum(\frac{\partial B_y}{\partial x}-\frac{\partial B_x}{\partial y})$. Mean current helicity $\overline{h_c}$ is calculated by $ \frac{1}{N} \sum |B_z (\frac{\partial B_y}{\partial x}-\frac{\partial B_x}{\partial y})|$. 
Mean free energy density {$\overline{\rho}$} is calculated by $\textstyle \frac{1}{N} \sum \frac{1}{8\pi}(B_o^2-B_p^2)$. 
All formulas {described above} are adapted from Table 3 in \cite{Bobra_2014}. 
\end{table*}

\clearpage

\acknowledgments{We thank our anonymous referee for his/her valuable comments that helped to improve this paper. We acknowledge the use of the data from {\it GOES}, from HMI and AIA instruments onboard {\it SDO}, and from EUVI, COR1, COR2 instruments onboard {\it STEREO}. 
L.L. is supported by the grants from the Open Project of CAS Key Laboratory of Geospace Environment. 
Y.W. is supported by the grants from NSFC (41574165 and 41774178). K.D. acknowledges the support by the Austrian Space Applications Programme of the Austrian Research Promotion Agency FFG (ASAP-11 4900217). M.T. acknowledges the support by the FFG/ASAP Programme under grant no. 859729 (SWAMI).  J.C. is supported  by NSFC through grants 41525015, 41774186. 

\begin{appendix}~\label{sec:appen}

\section{Quality of the NLFFF Extrapolation}~\label{subsec:quali_nlfff}
In {this} study, we reconstruct the coronal magnetic fields based on a NLFF 
assumption, which requires the model fields to meet {the} force-free and divergence-free condition. Following \citet{Wheatland_2000, rliu_2016, Lliu_2017}, we calculate two parameters: the fractional flux increase ($\langle|f_i|\rangle$) and the angle between the fields and the current density ($\theta$) in the extrapolated volume to check their qualities. The computing equations are: 
\begin{displaymath}\label{eq:fl}
\langle|f_i|\rangle=\frac{1}{n}\,\sum\limits_{i=1}^n\,\frac{|\nabla\cdot\bm{B}|_i\Delta\,V_i}{B_i\cdot\Delta\,S_i}
\end{displaymath}
\begin{displaymath}
\theta=\sin^{-1}\,\left(\left(\sum\limits_{i=1}^n\,\frac{|\bm{J}\times\bm{B}|_i}{B_i}\right)\,\Bigg/\,\sum\limits_{i=1}^n\,J_i \right)
\end{displaymath}
$n$ denotes the number of the cells in the calculation volume, $\bm{B}$ refers to the magnetic fields, $\bm{J}$ denotes the current density. subscript ``i'' refers to the $i_{th}$ cell, $\Delta V$ and $\Delta S$ refer to the volume and the surface area of each cell. 
See Figure~\ref{fig:nlfff_q} for the qualities 
of the model fields in the context. $\langle|f_i|\rangle$ are all well below 0.007 while $\theta$ are all well below 10$^{\circ}$, {confirming} that the model fields 
all meet {the} force-free and divergence-free condition. 

\begin{figure*}
\begin{center}
\includegraphics[width=0.9\hsize]{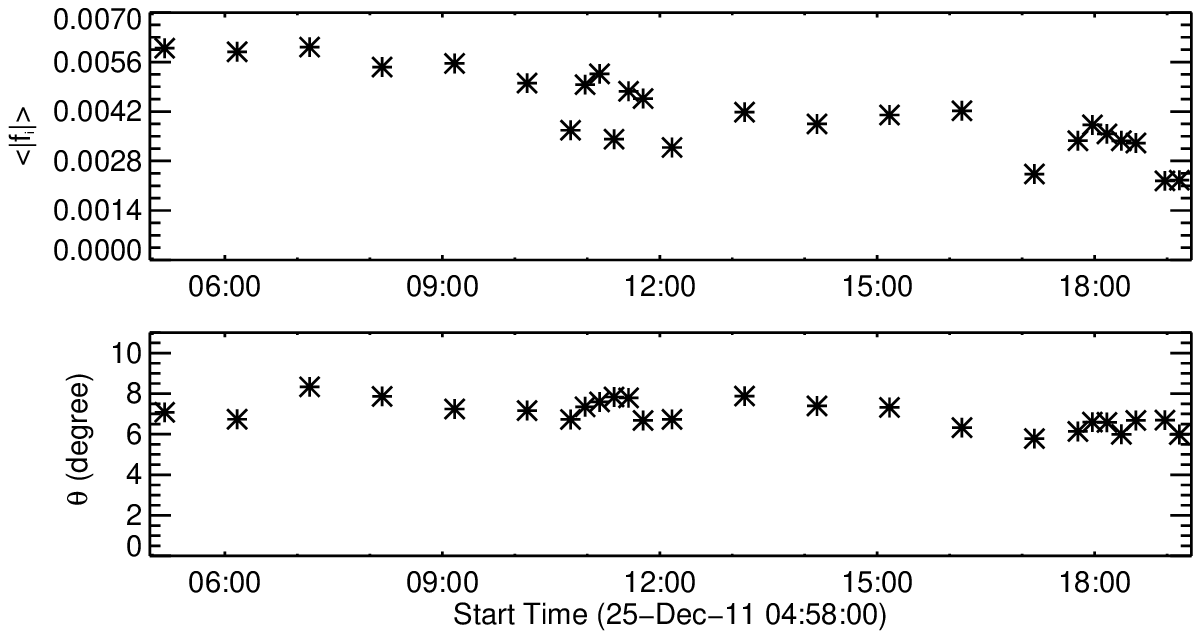}
\caption{Divergence-free and force-free parameters for the model fields in the context. The figure has the same time axis as Figure~\ref{fig:para_evo} (c). The upper panel displays the values of the divergence-free parameter $\langle|f_i|\rangle$. The lower panel shows the values of the force-free parameter $\theta$. }\label{fig:nlfff_q} 
\end{center}
\end{figure*}

\end{appendix}

\clearpage
\bibliographystyle{aasjournal}
\bibliography{con-eru}

\end{document}